\documentclass[10pt]{article}
\pdfoutput=1
\usepackage{amsmath}
\usepackage{graphicx}
\usepackage{epstopdf}
\usepackage{tikz} 
\usepackage{dcolumn}
\usepackage{bm}

\def \L {{\cal L}}

\renewcommand{\mathbf}[1]{\ensuremath{\boldsymbol{#1}}}

\title{ Relativistic Newtonian gravitation }

\begin{document}

\author { \\P.  Christillin, G. Morchio\\
Dipartimento di Fisica, \\
Universit\`a di Pisa\\ }

\maketitle

\begin {abstract}

\

The physical principles at the basis of an ``elementary derivation''
of the General Relativity (GR) effects in a static centrally symmetric field are
reexamined. We propose a theoretical framework in which all the GR results follow from
the EP, local SR and Newton law in intrinsic coordinates.

\end {abstract}

\vskip 1cm

    e-mail paolo.christillin@unipi.it, morchio@df.unipi.it

\section{Introduction}

The deviations from Newtonian gravity \cite{newton} in the centrally symmetric static case,
i.e. slowing of time, light deflection and precession of orbits
have played a fundamental role for the confirmation of General Relativity (GR)
\cite{einstein} . 

It is not clear to which extent the full GR theory is necessary for the 
prediction of such effects.
In particular, the possibility that the above three crucial tests
could be derived only or mainly from Special Relativity (SR) was first explored
in \cite{schiff}, but its conclusions have been criticized in
refs.\cite{schild} \cite{rind}. 
The problem of a derivation of the above effects
from a restricted number of assumptions has been reconsidered in
many other investigations \cite{cern}, \cite{gru},
\cite{vis},  \cite{padma},  \cite {kass} and  ref.s therein.  

While Schiff's analysis consists substantially in the attempt of a
direct intepretation of the metric of the popular Schwarzschild (Ss) solution,
the subsequent work has been directed to a similar interpretation of
the (less widespread) Painlev\'e-Gullstrand (PG) solution . 

For the Ss solution, as we shall see, the attempt to identify the correponding terms
as Special Relativity (SR) effects \cite{schiff} is only tenable for
time, whereas the factor given by SR for the radial coordinate is the inverse
of that.
The simplicity of the Ss solution, in particular the separation between
time and space effects in the invariant interval, both given by Lorentz factors, 
is in this sense misleading.



%
The PG metric
(originally interpreted as a
{\it physically different solution of the GR equations}) 
has also been obtained without recurring to
Eintein equations, by arguments which seemed
so "elementary" to be  considered as purely heuristic even by the authors
themselves. This sounds indeed a bit paradoxical since the ``elementary''
and GR approaches lead
to the same results, so that the {\it different}  ingredients of the first
cannot be discarded as fortuitous because they are not complicated enough.
This judgement is rather the result, in our opinion, of an uncomplete control
of the proposed theoretical framework,
related to the ambiguities in the identification of
the components of the metric tensor as physical objects
(see the discussion in Sect 4.3).

The purpose of the present paper is to reexamine the problem and clarify the ingredients
substantiating an elementary derivation of the GR effects in the static centrally
symmetric case.

While Schiff's considerations, based on the Ss solution,
try to reconstruct the GR results as
SR effects, in the approach based on the PG solution the Equivalence
Principle (EP) plays the most important role, the metric being the result
of an argument involving radial free fall trajectories,
provided by a velocity field $v(r)$
associated to gravity and common to all particles in free fall from infinity.

Free fall trajectories can be interpreted as defining 
a ``modified inertial law''. They have therefore nothing to do with accelerations in
Minkowski space and they should rather be regarded as transporting to all the space-time
points the metric of local MInkovski structure assumed   at infinity.

This amounts to use the EP in the form that, to the first order in the
displacements, the metric remains Minkowski for a free falling observer,
in the space and time variables defined by free falling clocks and rods.

The determination of the metric then depends on
the free fall law and on a parameter 
describing a possible space curvature, and 
the essential question is whether they are affected by the velocity of light.
One might indeed conceive a relativistic modification of Newton's law;
as we shall see, the result is sensitive to such corrections in the case of
Mercury's precession, which disproves their presence. 




On the basis of the above interpretation of free fall as a modified inertial law,
the basic idea is that only the Newton constant must be relevant for
the velocity of radial free fall and for the space geometry parameter \cite{gru},
in the intrinsic variables mentioned above.

This fixes the free fall law,
which holds at all distances and even provides a common description of the outer an inner
regions of the Schwarzschild black hole.
The velocity of light of course enters, but only
in the construction of the invariant interval, which is
at the basis of local relativistic physics.




We will show that indeed, in the static centrally symmetric case, the
ordinary Minkowski structure at space infinity and the Newton constant
uniquely identify radial trajectories and a metric on space-time, 
determined in physically constructed
(``intrinsic'') coordinates and coinciding with the one given by
the Painlev\'e-Gullstrand \cite {pain} \cite {gull} solution of the Einstein equations.

The presentation  will be elementary, starting from the modifications of
Newton's treatment required by an ``operational'' construction
of the space and time coordinates, which will be done on the basis of
the EP for radial free fall trajectories (Sect.2). The local Minkowski structure
will be determined in such coordinates in Sect.3. 
To better clarify the elementary character of our arguments,
in Sects.4 and 5 the predictions of the main GR effects will
be derived directly from our approach.

The results will be compared to other approaches,
and a discussion of 
the Sagnac effect along similar lines will follow.

\

\section {Newtonian time and the Equivalence principle} 

\

a)  {\bf absolute time from the EP} 

\

Newtonian space-time consists of Euclidean space and ``absolute'' time,
a notion which clearly conflicts with the basics of Special Relativity.
This justifies  the general consensus about the fact that  the inclusion
of SR into Newton's theory be forbidden from the beginning,
leaving as the only solution a complete reformulation of the entire problem of gravity,
which is in fact the case of the General Relativity approach.

However, it is our aim to show that there is a simple and direct way to give a meaning
to ``relativistic corrections to Newton gravity'',  on the basis of the Equivalence Principle
and Special Relativity.  This can be done in a {\it substantially unique way},
and reproduces  all the General Relativity results for static central
gravitational fields.

The basic form of the EP is that feathers and lead balls
experience the same gravitational effects.
All objects acquire the same free fall velocity: 
their {\it  inertial and gravitational masses are equal}.

In general, the purpose of the EP is to use frames associated
to free falling observers to ``eliminate'' gravity locally,
to the first order in the space-time displacements from a given point.

In the following, we will use the EP to include gravity effects in the 
 Newtonian notion of time. 
{\it We assume central symmetry and the validity at large distances of the
Newtonian description: space is Euclidean and time satisfies,
in that limit, all the synchronization properties defining  Minkowski frames in the 
absence of gravity.  We also assume stationarity with respect to the time at infinity}.
Then, the EP, which asserts that gravity effects are not
felt by free falling observers, suggests to \emph{define} time by clocks in free
fall from infinity.

We assume therefore that clocks can be arranged to fall freely from infinity,
starting at all times, along radial trajectories, with zero initial velocity; they
provide a  unique notion of time, defined for all space-time points.
We will adopt such a notion of time as the {\bf ``EP absolute time "  $\mathbf t$} with
a similar role as Newton's absolute time.

Let us also remark that, by the above stationarity assumption, the time needed for
free falling clocks to reach a given point in space
from infinity is independent from the starting time and that therefore in this sense
time intervals at any space point ``coincide with time intervals at $\infty$''. 

We assume invariance of all the physical laws with respect
to the translations of the EP absolute time.

Notice that \emph{no velocity parameter appears in such a construction}, due to
the null velocity of the clocks at infinity. Other possible constructions,
with a non-zero velocity at infinity, would require the use of SR to account for
the initial motion near infinity, preventing a clear separation of roles between SR and EP. 

Clearly, the introduction of the above notion of time has important 
consequences, even on the description of space alone, since the very 
identification of the space variables and of space geometry 
concerns, by definition, space-time points at the same time.
This affects in particular the notion of space distances, which will
be defined as measured by sequences of small rods, with their 
ends taken at the same time.

The same construction can be performed for radial trajectories \emph{reaching}
infinity (at time $+ \infty$)  with zero velocity.
Our results will be independent
of the choice between the corresponding (alternative) notions of time.
To be definite we will consider in the following the case of infalling
velocities  i.e.  $v(r) < 0$.


\

b)  {\bf Newton laws } 

\

Adopting the above reformulation for time, we now endorse the Newton
principles of gravitation, for a centrally symmetric static gravitational 
field:
\

1) {\it space is assumed to be euclidean}. This amounts, 
due to central symmetry, to the euclidean relation between radial and angular
distances 
\begin{equation} \label{distan}
 \oint dl = 2 \pi r  
\end{equation}

2) {\it radial free fall is asserted to be given by the Newtonian 
velocity law}
\begin{equation} \label{velo}
v^2(r)= 2GM/r = - 2 \Phi(r)
\end{equation}

As well known, since the same law applies to all bodies, eq.(\ref{velo} )
includes, for the case of radial free fall from infinity,
the basic form of the EP , i.e. the uniqueness of 
free fall trajectories for given initial position and velocity.

We emphasize that all the above notions refer to \emph{measured} space
and time intervals;
as a consequence of assumption 1),
space distances are given by the euclidean expression
in Cartesian
coordinates $x$;
the velocity in eq.(\ref{velo} ) is defined by the above measurements of space distances and
by time intervals given by free falling clocks.

Opposite to the ordinary GR point of view, we do not start from coordinate independence, but
rather identify coordinates allowing for a description in the spirit of Newton gravitation.
We have therefore 

\noindent
1) \emph {chosen} to discuss the property of space and time in presence of
gravity in ``intrinsic coordinates'', obtained in terms of Euclidean coordinates near
infinity, extended by using times and distances measured by radially free falling observers.

\noindent
2) \emph {assumed} Euclidean space and Newton's free fall law in such coordinates

The velocity of light does not appear in the above considerations: only the Newton
constant $G$ and the mass $M$  enter in the above description of space and time. 
On such a basis
Eqs.(\ref{distan}),(\ref{velo}) are forced
by dimensional analysis and space flatness  at infinity,  which determines in particular 
the value $2 \pi$ in eq.(\ref{distan}).

\

c) { \bf Free Falling Frames and the EP} 

\

So far  the EP has only been used to derive a notion of time, in which
Newton laws have been expressed. Let us now formulate the { \it complete}  
EP. To this purpose, the essential step is to introduce, around each space-time point,
local variables associated to Free Falling Frames.

Observers falling along radii starting at infinity (at time minus infinity) with zero velocity
employ, around  a trajectory $X (t)$,  time intervals measured by free falling clocks,
and space distances from $X(t)$,  measured by (small) rods with ends
at the same time and therefore given by euclidean expressions in $ x- X(t)$.
Using $dX/dt = v(X(t))$, $v(r)$ given by Eq.(\ref{velo} ),
the corresponding differentials at a space-time point $r_0, t_0$ are
\begin{equation}  \label{pseudogal}
\begin{split}
  & dt' = dt  \\
  & dr' = dr - v(r_0)\, dt \\
  & dx'_\perp = dx_\perp
 \end{split}
 \end{equation}
 with $ dx'_\perp $ the space  displacements in the directions orthogonal
to $r$, given by the differential $ dx_\perp $ of local Newton cartesian coordinates
orthogonal to the radius.

\begin{figure}
\includegraphics[width=0.52\columnwidth, clip]{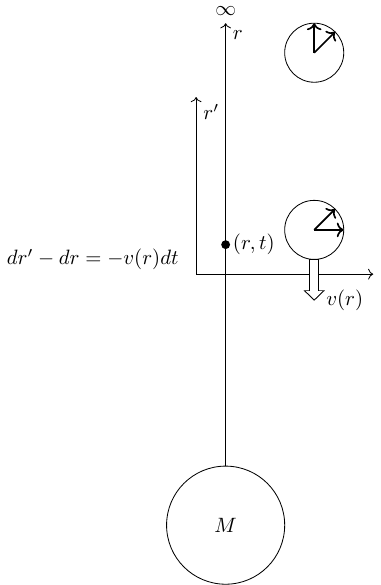}
\caption{  The clocks of the infinitesimal Equivalence Principle Inertial Frames
  (EPIFs), starting at any time,  associate to each point the  time  
  $t'= t$. The relativistic space-time effects of gravitation are
  determined  by the infinitesimal invariant Minkowski 
interval $ ds^2 = c^2 dt'^2 - dr'^2$ in the EPIFs coordinates , which 
gives the metric in the  global coordinates  $(r,t)$ through Eqs (3).} 


\end{figure}

Even if Eqs.(\ref{pseudogal})  have the form of Galilei transformations, they have a very
different nature, since they describe ``small'' (infinitesimal) displacements in
Free Falling Frames, on the l.h.s.,  in terms of global coordinates in the r.h.s.. 

{\it It is a fundamental fact that the above relations are  \underline {not} given by
  Lorentz transformations.  They have the form of Galilei transformations
  because they arise from the use of a common, ``absolute'', notion of time.}

Eqs.(\ref{pseudogal}) should not be interpreted therefore as a low velocity approximation of
Lorentz transformations (as in \cite {vis}).
They  are  exact in our approach, for all values of the free fall velocity.
By the above derivation, they hold independently of  eq.(\ref{velo}), which only
fixes the value of $v$; morever, only the third equation should be modified
(by an $r$ dependent factor) in the absence of the Euclidean relation eq.(\ref{distan}).

In fact, in the present Section we are only discussing the modifications to the
{\it inertia principle}, which holds both in relativistic and non relativistic
physics and has little to do with the velocity of light, which was in fact never mentioned
in the above discussion.

Clearly, the use of a Lorentz transformation in
Eq.(\ref{pseudogal} ) would  lead to a trivial local Minkowski structure,
excluding gravity effects on clocks and light deflection.

The differentials $dx'$ and $dt'$ representing the description of space and time 
in Free Falling Frames, {\it to the first order in the displacements} from
a free fall trajectory,  are the  
substitute of coordinates satisfying the inertia principle. They
will be denoted as (infinitesimal) Equivalence Principle  Inertial Frames (EPIFs)
and are the object of the following form of the \emph{Einstein's Equivalence Principle}: 

\emph {All the physical laws which can be written, in the absence 
of gravity, in inertial frames in terms of local variables 
and their first order variations around each point, hold true 
in the presence of gravity in terms of the same variables
in EPIFs}.  

It is important to notice that the \emph{differentials} defining EPIFs
\emph {do not in general define coordinates, even locally}.
In fact the differential form 
$$   dr' = dr - v(r)\, dt  \  $$
 \emph {is not integrable}, unless
the velocity field $v(r) $ is constant  (the trivial inertial case), since
$$ \partial v /\partial r = - \partial  \; 1/ \partial t = 0 $$
is precisely its integrability condition.

{\it On the contrary for the time variable the restriction to the infinitesimal
  interval is not essential and in fact $t=t'$ is  the ``gravity free''
  time measured by clocks on EPIFs, which does not suffer from the limitations
  produced by gravity on the space variables. }

Clearly, even if the formulation of the EP only uses EPIFs,
its implications crucially depend on the
relation between the above differentials at different
points, given by Eqs.(\ref{pseudogal}).
In other terms, the introduction of global coordinates  and the expression of the EPIF differentials in terms of them is an essential step
for an effective use of the EP.

\

\section {Relativistic Physics}

\subsection{Basic gravitational effects}\label{basic}

\

a) {\bf Newton's mechanics from the principle of least action in EPIFs}

\

Let us first show how the use of the  EP reproduces classical
non-relativistic mechanics  for a particle in the gravitational field of a mass $M$.

Classical Mechanics can  be indeed formulated in terms of the
principle of least action in  inertial frames.
The implementation of the EP in the non relativistic free Lagrangian
$ {\cal L}  =  m/2 \;  \dot x^2 $ is immediate.
Following  Eqs. (\ref {pseudogal}), it is enough to express the velocity in the local EPIF  

$$ \dot r     \rightarrow \dot r - v(r)$$   

{\it The principle of least action in EPIFs for a free
  particle in the presence of gravity} is thus given by 

\begin{equation}
  \label{NRSP}
\delta   \int  dt  \;(m/2\; [(\dot r-v(r))^2 + r^2 (\dot \theta)^2]) = 0 \, .
\end{equation}

The Lagrange equations yield for the radial coordinate

\begin{equation} \label{NewR}
\begin{split}
d/ dt (\dot r-v(r)) + (\dot r-v(r))dv/dr - r (\dot \theta)^2  =\\
=  \ddot   r - d/dr (1/2v^2(r)) - r (\dot \theta)^2  = 0 \, . 
\end{split}
\end{equation}
Since $ 1/2 \;  v^2(r) = GM/r$,
the Newton's radial equation of motion is obtained, and the same holds
also for the angular variables.
A constant $v$ would only amount to a change of inertial frame.

\

b) {\bf The invariant Minkowski interval} 

\

Relativistic physics is governed by the ``infinitesimal'' invariant interval.
By the EP,  the invariant interval has the standard Minkowski form in EPIFs, in which
the ordinary inertial frame laws hold and light propagates isotropically and
always with velocity $c$, to first order in space-time displacements:
\begin{equation}  \label{Minkint}
 ds^2 = c^2 dt'^2 - dr'^2 - dx'^2_\perp  \, .
\end{equation} 

The local (first order) validity of the principles of SR 
implies that $ds^2$ {\it is  still given by the same expression}, Eq.(\ref{Minkint}),
in the coordinates employed by any observer around the given space-time point,
to the first order, independently of his motion.
All the SR results hold  locally,
{\it  for all observers (on arbitrary trajectories) to first order in the
coordinates defined by their clocks and rods},
with the Minkowski interval given by the ordinary expression.

The above expression of the EPIF differentials in terms of globally defined variables
allows to write the Minkowski intervals, all of the same form in their EPIF variables 
around different points, in global coordinates:
\begin{equation}
  \begin{split} \label{Minkint2}
      ds^2 = c^2 \, dt^2 - (dr - v(r) \, dt)^2 - dx_\perp^2 = \\
    = c^2(1 - v^2(r)/c^2) \,dt^2 + 2v(r) \,  dt \; dr - dr^2 - dx_\perp^2 
\end{split}
\end{equation}

Eq.(\ref{Minkint2}) gives nothing else than
{\it the Painlev\'{e} \cite{pain} - Gullstrand \cite{gull} metric (P-G),
  a solution of the GR equations in a central field}, obtained here
(as a solution of no whatsoever equation other  than Newton's law) {\it  on the pure basis of
  Euclidean space and absolute ``free fall'' time. }
Even if built on Euclidean space and absolute time, it represents
a non-Minkowskian space-time, due to the  crossed term.
The P-G metric is equivalent to the Schwarzschild \cite{ss} metric , the relation being
given by a change of the time variable  (see SEction4.2).

Clearly, in our  approach, SR enters at a different stage with respect to the 
description of  radial free fall. The latter is given by the Newton free fall
velocity in global coordinates; the velocity of light  enters in the local relativistic
structure of space time, which is trivial in EPIFs and globally determined by
Eqs.(\ref{Minkint})(\ref{pseudogal}).



\

\

\

c)   {\bf Clocks ticking and red shift } 

\

How time flows  in a gravitational field for observers at rest, in  the above (P-G)
coordinates, is immediately got from the P-G metric.  Actually, the notion  of rest is independent of coordinate transformations preserving
stationarity of the metric tensor. 
By setting $dr =0$,
\begin{equation}
d\tau^2 =  (1 - v^2(r)/c^2) \, dt^2  = (1 -  \epsilon(r)) \, dt^2
\end{equation}
{\it relates the (P-G) Newtonian free fall absolute time $t$
  to the relativistic invariant  interval $d\tau$ measured by observers at rest in
  the P-G coordinates, thus defining their proper time.}

The parameter 
\begin{equation}
\epsilon(r) \equiv  2 GM/rc^2 =  v^2(r)/c^2 \equiv - 2  \Phi(r)/c^2  
\end{equation}
does not appear for non relativistic mechanics and
enters in our approach only through the invariant interval.


By time translation invariance and linearity of propagation,
the frequency of light
propagating from infinity remains constant in (the above, P-G) time, 
and therefore frequencies observed by observers at rest are given by (the
inverse of) the above relation.

Because  the velocity of light remains the same for all observers,
{\it this can also expressed in terms of wavelengths},
  relating the one at $\infty$, $\lambda_\infty$ to the one at $r$ $\lambda_{r}$, 
 $$ \lambda_{r} = ( 1 - \epsilon (r))^{1/2} \, \lambda_\infty
  \simeq ( 1 - \epsilon(r)/2) \, \lambda_\infty  $$
  For small radial distances $h$,
  $$ \Delta \omega / \omega \simeq   \epsilon(r)/2 \; ( h/r)  $$
i.e. the well known red shift. For moving sources,
one has to add the usual Doppler effect.

\

d){ \bf Light cones}

\

Light velocity is obtained by setting to zero the invariant interval  Eq.(\ref {Minkint2}).
The velocity in directions orthogonal to the radius takes the value 
\begin{equation}
   c_\perp = r d\theta /dt = c \, (1- v^2/c^2)^{1/2} \, .
\end{equation}
Along the radius, the velocity is 
\begin{equation}
c_r = dr /dt  =  \pm c + v(r) \, ,
\end{equation}
$ v = v(r)$, given by eq.(\ref{velo}) ($ v<0 $).
 {\it Both equations directly follow from eqs.(\ref{pseudogal})
 by ordinary (Galilean) vector composition of the (isotropic) velocity ${\bf c}$
  in EPIFs  with the EPIF velocity $v$},
\begin{equation} \label{vellight}
  \mathbf{c}_{PG} = \mathbf{c} + \mathbf{v}(r) \, .
\end{equation}

\

\begin{figure}
\includegraphics[width=0.82\columnwidth, clip]{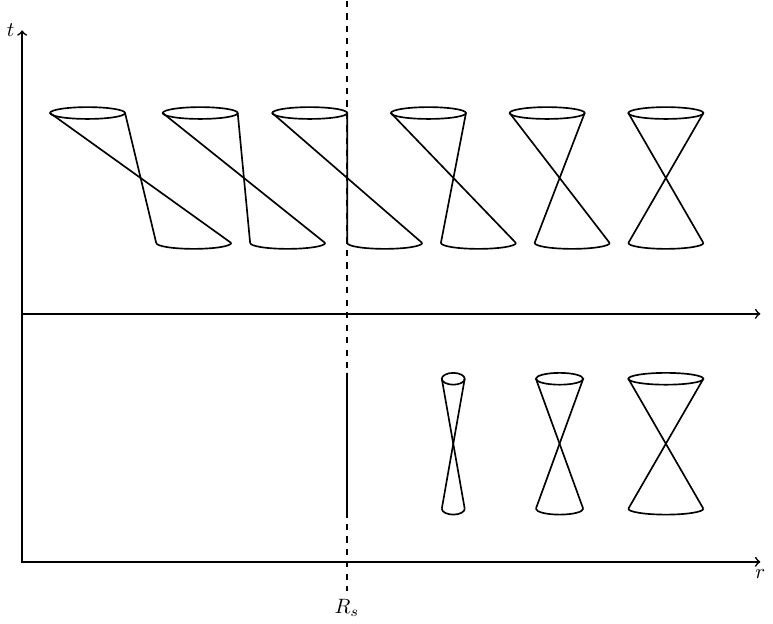}
\caption{ Light cones as a function of $r$ in the  Painlev\'e-Gullstrand
  and Schwarzschild metric respectively.  A singularity at $R_{S}$ only arises for
  the second metric; in the P-G coordinates , because of the distorted
  light cone, nothing can simply get out of the hypothetical black hole.}
  
\end{figure}

\

For comparison, we recall the results for the Schwarzschild metric 
\begin{equation}
ds^2 =(1-v^2/c^2)  c^2  dt_{S} ^2 - dr_{S} ^2/(1-v^2/c^2) -  r^2 d\Omega^2 \, .
\end{equation}
There the radial velocity is 
\begin{equation}
dr/dt  = \pm  c \; (1 - v^2/c^2)  
\end{equation}
whereas the tangential one is the same as in the P-G coordinates.

Thus the Schwarzschild light cones shrink for decreasing distances down to a pseudo
singularity at $R_{S} = 2MG/c^2$. In the P-G coordinates, they rotate, without 
any singularity.    
If one {\it believes} in the validity of  the extrapolation
of Newtonian dynamics to such extremes (to be discussed later),
a physical effect emerges from the SR constraints in EPIFs.

Indeed,
when $v(r) < - c$ light cannot propagate outwards
for positive times.

This happens below $R_{S}$, where the P-G metric describes a black hole.
Reversing the sign of time is equivalent to reversing the free fall velocity
in the above construction. In this case, light cannot propagate inwards
(since this time $v(r)  - c  >0$), for positive times and the corresponding
P-G metric describes a "white hole". 


Notice that the result only depends on the free fall velocity
(in intrinsic coordinates) exceeding $c$ at some radius, and has therefore
nothing to do with any ``interior dynamics'' below such a radius.


\

e) {\bf Relativistic Mechanics}

\

The formulation of relativistic mechanics is immediate via the EP,
which only amounts to substitute in the variational principle   the ordinary Minkowski
intervals  with the Minkovski
intervals in EPIFs. The corresponding action is therefore obtained,
as in the non-relativistic case, by the
substitution   $ \dot r \to \dot r-v(r) $

\begin{equation} \label{Lagr}
  \delta {\it  A} = {dt}  = \delta   \int  {ds} = \delta   \int   {\cal L}  \;      
  \delta   \;  \int  dt \; (m c^2)
  \sqrt{  (1- 1/c^2 [(\dot r-v(r))^2 + r^2 (\dot \theta)^2])}  =  0
\end{equation} 

The equation of motion are given by the corresponding Euler Lagrange equations.
They are equivalent to the geodesic equations in the metric
defined by the above invariant interval, i.e. in the P-G metric.


Both solve in fact the same variational problem, the ordinary geodesic
equations being obtained from a parametrization of trajectories with
the proper time
and the substitution of the Lagrangian
with its square (which is allowed since the Lagrangian associated
to the proper time parametrization takes the value $1$ on the 
solution of the stationarity problem).
Let us also mention that the use of proper times is inappropriate
in the many body case.

\

\subsection{Light deflection}\label{light}

\

Since our invariant interval coincides with the one of the Painlev\`e-Gullstrand
solution of the Einstein equations and the principle of
stationary action amounts to geodesic motion in the corresponding metric,
all the results of GR  for the dynamics of particles and light follow.

We show below that these results can also be derived directly in a rather elementary way.

 

The problem of light bending has been paramount in assessing
the view of space distortion associated to the Schwarzschild solution.
 Indeed, classically, (see e.g. \cite {berk}) it is well known that Newtonian mechanics
 can account for light deflection, at variance however by a factor of 2 from the
 GR result and from experimental data.
 This is explained in Schwarzschild coordinates by saying that Newton
 just reproduces the  time part ($g_{00}$) of the metric tensor and that the
 space part  $g_{ii}$  is the new fundamental contribution of GR.

Let us first recall the classical treatment and then discuss the contributions  
arising from the EP. We restrict to first order in the (relativistic) parameter $\epsilon(R)$,
which completely covers the experimental situation.

Take a  luminous  ray grazing the sun, coming from infinity and calculate the
{\bf light  deflection} observed at large distances
(on the earth, practically at infinity) .

\

{\bf a) Newtonian light deflection}

\

To first order, the bending angle of light associated to the unperturbed trajectory
$(x = ct, y=R)$, is given by $ \theta(x) = - c_y/c $, $\ c_y$
the $y$ component of the light velocity,
$c$ the  unperturbed velocity.   If light is assumed to accelerate
according to Newton's law 
\begin{equation}
  \label{Newtontheta}
  \theta (x) \simeq -  c_{y} (x) / c =
  \int_{-\infty}^t \frac{\partial }{\partial R}  \Phi(ct' , R) \;  dt' / c =
  \int_{-\infty}^x \frac{\partial }{\partial R}  \Phi(x' , R) \;  dx' / c^2 \, ,
\end{equation}
$\Phi$ the Newton potential, eq.(2).
The deflection angle is then obtained by integration over the whole $x$ axis in
terms of the {\it relativistic}  weak field parameter 
$  2 GM/ c^2 R \,  =\epsilon(R)$
as 
\begin{equation} \label{Newthinf}
\theta =  \theta(\infty)  = \epsilon(R)
\end{equation}

\

{\bf b) Wave fronts and light velocity }

\

In order to discuss light deflection as a refraction effect
produced by position dependent velocities, let us see how it can be obtained in general,
for small angles, in terms of wave fronts. 
The Newtonian result will be shown to follow from the light velocity given
by the pure time component of the Schwarzschild metric, while the full GR result
will follow from the above (very elementary) P-G light velocity.

We  will consider the propagation of light in a first approximation
along straight lines (see Fig.2)) at different heights , with velocity $ dx/dt = c(x,y)$
and calculate the orientation of wave fronts, at $y \simeq R$.

Since, as motivated before,  light frequency $\omega$ remains constant 
in the P-G time $t$ (and also for the Schwarzschild time, see below),
the phase $\varphi$ of the wave changes with the time it takes a wavefront
to travel in the $x$ direction  
$$ d \varphi  = dx / \lambda(x) = \omega \, dx / c_x(x,y) \,  , $$
$c_x $ the velocity of propagation along the $x$ axis.
Thus, to first order, the difference of the wavefronts at different heights 
\begin{equation}\label{dphi} 
  \frac{\partial} {\partial y} \, \varphi (x,y)    \simeq
  - \int_{-\infty}^x \frac{\partial} {\partial y} \, c_x(x',y)^{-1}   \; \omega \, dx'
\end{equation}
determines the bending. 
With the same sign convention as above the bending angle $ \theta$ of the
wavefront at $ y \simeq R$ is given by

\begin{equation}\label{wavetheta} 
  \theta (x) 
  = -  \frac {\partial } {\partial R} \,  \varphi(x,R)  \; \lambda(x) \simeq
   - \int_{-\infty}^x    (\frac{\partial } {\partial R} \, c_x(x',R)^{-1})  \, c_x(x,R) \, dx'
\end{equation}
and the deflection angle is $\theta =  \theta(\infty)$.

\

{\bf c) Schwarzschild }

\

The modifications of the light velocity given by the
``pure time component'' of the Schwarzschild metric,
$ ds^2 = c^2 \, (1  - v(r)^2/c^2) \, dt^2 - dx^2 $, are
independent of the direction and given by
\begin{equation}
  c(x,y)=  c \, (1 - v^2(r) /c^2)^{1/2}  \simeq    c  (1 -\epsilon/2) 
\end{equation}
and therefore Eq.(\ref{wavetheta}) gives the deflection angle 
\begin{equation}
  \theta    \simeq   \int_{-\infty}^\infty 
   \frac{\partial } {\partial R} \, \Phi(x,R) / c^2  \, dx \, ,
\end{equation}
which coincides with the Newtonian expression, eqs.(\ref{Newtontheta})(\ref{Newthinf}).
The Newtonian equation of motion is in fact equivalent to the stationarity
principle for the optical length in such a metric.

For the complete Schwarzschild metric, the velocity of propagation
of light along the $x$ axis, with $y = R = r \cos \alpha$, is given by
$$
 c^2  dx^2 ((1 - v^2/c^2)^{-1} \sin^2 \alpha  + \cos^2 \alpha) = (1 - v^2/c^2) dt^2
$$
which gives  
\begin{equation}
  c_x(x,R)^{-1}    \simeq   c^{-1} \, (1 + \epsilon(r) \, (1 - R^2/(x^2+R^2)/2) ) \, .
\end{equation}
The integral of the last term is finite and independent of $R$, so that
the result  changes by the well known  factor  of $2$.

\

{\bf d) deflection from EPIF Galilean light velocity composition}

\

In our approach light  velocity is given by the Galilei formula,
eq.(\ref{vellight}).

\begin{figure}
\includegraphics[width=0.82\columnwidth, clip]{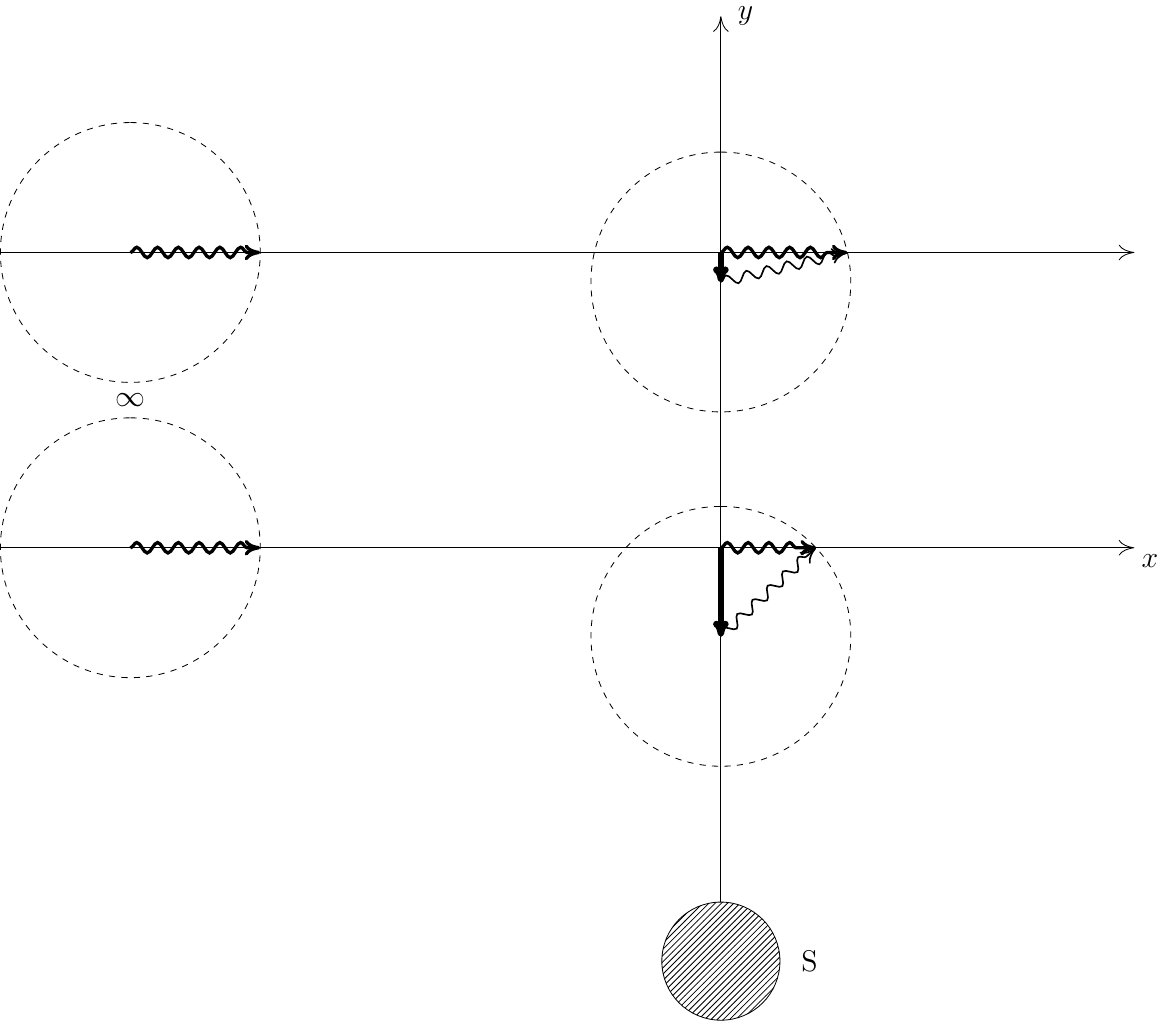}
\caption{ Spherically symmetric  light propagation in the EPIFs is vectorially composed with the free
fall velocity ${ \bf v} (r)$ to yield a resultant $c_x$ along the unperturbed trajectory.
The phase variation and hence the bending of
the wave front as a mirage effect comes from the dependence of the $x$ velocity
$c_x$ on the height $y$.}
  
\end{figure}

By imposing to first order its propagation along the $x$ axis, the $y$ components
cancel in eq.(\ref{vellight}) and therefore the $x$component is given by 
$$    c_x =  (c^2 - v_y^2(r))^{1/2} + v_x(r) $$

Since $v(r)/c$ is if of order $\epsilon(R)^{1/2}$, to first order in $\epsilon$  
\begin{equation}\label{cmuPG}    
c_x^{-1} =  c^{-1} \, ((1 - v_y^2(r)/c^2)^{1/2} + v_x(r)/c)^{-1} \simeq 
\end{equation}
$$ 
c^{-1} \, (1 + v_y^2/2c^2 - v_x(r)/c + v_x^2(r)/c^2) = 
  c^{-1} \, (1 - v_y^2/2c^2 - v_x(r)/c + \epsilon(r) )    $$
The second term is proportional to $\Phi(r) R^2/(x^2+R^2)$ and its integral
is independent of $R$, as above. The third is antisymmetric in $x$ and its
integral vanishes;  the last term gives the GR result,
$$  \theta =  \theta(\infty)  =  2 \;   \epsilon(R) $$

The factor 2 has emerged from the {\it second order} expansion of the
inverse of the velocity in the parameter $\sqrt{\epsilon(r)} $.

Notice that in the P-G coordinates light velocity is different
on its way towards and away from  the source of gravitation,
due to the linear dependence on the
free fall velocity. As we have seen, light deflection arises as a second order
effect in the free fall velocity.
Linear terms are present in the deflection angles
at finite distances from the source; they 
depends crucially on the notion of simultaneity implicit in the definition of
wave fronts, which is  different in different coordinate systems.

\

\subsection {Perihelion precession}

\

Let us show how the perihelion precession can be calculated directly,
{\bf for motion close to a circular orbit and to first order in $\epsilon$},
from the above equations of motion.

The relativistic Lagrangian is given by eq.(\ref{Lagr})
The angular equation of motion is
\begin{equation}
 d/dt \ \partial {\cal L} / \partial \dot \theta  =  
 - d/dt \, ({\cal L}^{-1} r^2 \, \dot \theta) = 0 \, , 
\end{equation}
i.e. ,
\begin{equation} \label {angmom}
L \equiv r^2 \, d\theta/dt \, {\cal L}^{-1} =  r^2 \, d\theta/ds = const 
\end{equation}
The radial motion is given by:
\begin{equation} \label {radial}
  d/dt \, \partial {\cal L} / \partial \dot r - \partial {\cal L} / \partial r  = 0 \, .
\end{equation}
By using the proper time, here dentoted by $s$, $ds/dt =  {\cal L} $, we obtain
$$  d^2r/ds^2   =   v (r) (d/ds \, {\cal L}^{-1})   
   - {\cal L} ^{-2} d \Phi/dr +  L^2/r^3 \,  = 0  \, . $$
This equation is equivalent to the Schwarzschild equation
\begin{equation} \label {Ss}
d^2r/ds^2 = -  d\Phi/dr  + L^2/r^3 (1 - 3/2 \epsilon )
\end{equation}
since they solve the same variational problem in the same variables,
and therefore implies the GR result for the perihelion.
Even if the P-G radial equation is more involved (a dependence on $r$ and $dr/ds$
being hidden in the terms involving the Lagrangian), a calculation of the
precession effect  in the above approximations is straightforward.

Let us derive it directly from the equations of motion in
our (P-G) time variable. The radial equation (\ref{radial}) reads
$$ - d/dt \, ( \L^{-1} \, \dot r)  + (d/dt \,  \L^{-1}) \, v(r) -  \L^{-1} \, d\Phi/dr
+ \L \, L^2/r^3  = 0 \, . $$
 Multiplying by $\L$,
\begin{equation} \label{radialt} 
  d^2r/dt^2  = - (\dot r - v ) \, \L \,  d/dt \, \L^{-1} \,   
 -  d\Phi/dr + \L^2 \, L^2/r^3
\end{equation}
Circular orbits are given by
\begin{equation} \label{circorb}
 - d\Phi/dr + \L^2 \, L^2/r^3  = 0
 \end{equation}
and their  frequency is  
\begin{equation}
  \label{omegatheta}
  \omega_\theta^2
  \equiv \dot\theta^2 =  \L^2 \, L^2/r^4  = GM/r^3 \, ,
\end{equation}  
the same as in the Newton case.

Eq.(\ref{circorb}) differs from  Newton's equation by  the term 
$$\L^2 =  1 + 3 \epsilon$$

For small oscillations  around a circular orbit
 the term  $\dot r \, \L \, \dot \L^{-1}$,
quadratic in $\dot r$,  can be dropped; using $v ^2 = - 2 \Phi$, 
$$ \L \sim  1 + (\Phi  + v \dot r - L^2/ 2 r^2)/ c^2 $$
and the circular orbit constraint,
the terms linear in $\dot r$ of Eq.(\ref{radialt}) are readily seen to cancel,
corresponding to the absence of damping.
As a result, the only contribution of the first term in the r.h.s of Eq.(\ref{radialt}) is
$$ - v^2/c^2 \, d^2r/dt^2  =  - \epsilon  \, d^2r/dt^2 \, .$$

Eq.(\ref{radialt}) therefore reduces to
\begin{equation}
  (1 + \epsilon) \,  d^2r/dt^2  =  -  d\Phi/dr +
\L^2 \, L^2/r^3 = -  d\Phi/dr +
  (1 - \epsilon  - L^2/ r^2 c^2)
  \, L^2/r^3 \, . 
 \end{equation}
The  frequency for circular orbits  is thus given by
$$ \omega^2_r (1 + \epsilon) =  d^2\Phi/dr^2 -  \L^2  \, d/dr \, L^2/r^3
+ L^2/r^3  \, d/dr \,  (\epsilon(r)  + L^2/ r^2 c^2)   $$
$$ \sim  \omega_\theta^2 + d\Phi/dr \, 2 d\epsilon/dr  $$
so that 
$$ \omega^2_r \sim \omega_\theta^2 ( 1 - 3  \epsilon )  = GM/r^3 \  (1 -3  \epsilon) $$
i.e. 
$$ \omega_{r}/  \omega_\theta \simeq  ( 1 - 3/2  \; \epsilon ) \,  $$

and the precession angle is therefore
\begin{equation}
\frac{\Delta \phi}{2 \pi}  =   \frac{3}{2} \; \epsilon = 3  \frac{ GM}{rc^2}
\label{perih}  
\end{equation}

Since $ GM/rc^2 = v^2/c^2$,
$v$ the Newtonian velocity for circular orbits,
the result can be interpreted as a correction given by the relativistic equation of
motion, in a gravity field which is  described by
Newton law (in intrinsic coordinates).

A ``relativistic'' ($O(GM/rc^2 $) correction
to Newton law would result in an additional term,
of the same order, in Eq. (\ref{perih}).
On the contrary, such a correction would give a {\emph second order}
contribution to time ticking and light bending; therefore, 
the perihelion precession, usually interpreted as \underline{the} test of GR
(see e.g. Schiff  \cite{schiff}), can be equally seen as the
test of Newton law in intrinsic coordinates.

Notice, in connection to MOND \cite{mond}, that the relativistic corrections appearing
in the above equations have nothing to do with effects of order $O(v^2/r)$.
In other words comparable velocities (e.g. Earth and orbiting HI lines),
even at very different radii, have the same sort of relativistic corrections
(with negligible effect in the second case).

\

\

\section{Dynamics, metrics, observables  and all that}

\

\subsection  {The Newtonian fall velocity and the mass} 

\

So far our treatment has relied on a somewhat abstract framework,
assuming that {\emph the free fall velocity is given by Newton's law}.
Here  we want to ascertain to which extent this assumption follows
from dynamical considerations in Minkovski space. 

To start with, assuming that energy is the source of gravitation,
the mass in the potential term in Newton formula
should be corrected both by the self energy and by the kinetic term.  
  
Energy conservation for our free falling particle would thus read
 \begin{equation}\label {fundamental1}
m_{0}v^2/2  =  \frac{GM m_{0}}{r} (1-GM/c^2r  +v^2/2c^2) 
\end{equation}
or 
\begin{equation}\label {fundamental2}
v^2/2  -  \frac{GM }{r} =   \frac{GM }{c^2r}(v^2/2-GM/r) 
\end{equation}
whence 

\begin{equation} \label{fund}
v^2 = 2 GM/ r \, 
\end{equation}
The result only uses conservation of energy. 
{\it This puts Newton's law on a somewhat safer ground in the sense that
the above energy corrections cancel out.}
It is paramount to underline that {\it 
  the self energy correction to the mass, which embodies the fact that
  the graviton is itself a source of gravity, a relativistic  effect, is
  cancelled for radial trajectories  by the relativistic kinetic
  corrections of the gravitational mass.}

This has to  be contrasted with what happens in the PPN parametrization
(see e.g. \cite{damour}),
where the non linearity of Einstein equations
appears in the form (destitute of any measurement prescription)

$$ {\it h _{00}} =   2GM/c^2x (1- GM/c^2x)  $$

Thus one concludes that, remarkably, the non linearity of gravitation may depend
on the formulation. 
  
We also notice that the Newtonian  $1/r^2$ form  of the force is crucial in canceling
possible contributions from external masses, within a reasonable schematization of
the outer world as a homogeneous sphere. 

One might inquire about other relativistic corrections to the preceding expression.  

A possible relativistic extension of Eq.(\ref{fund}) is
\begin{equation}\label {fundament}
  \frac{m_{0}}{\sqrt{1 -v^2/c^2)}}  - \frac{GM m_{0}}{ c^2 r}
  (  1/\sqrt{1 -v^2/c^2)} -GM/c^2r) =   m_{0} \, .
\end{equation}
The terms in brackets sum up to 1 as in Eq. ( 33-35) so that the only possible correction is given by the l.h.s.  
It is important to notice that {\it the inertial mass
  in the l.h.s. cannot be gravitationally corrected}  by an
additional factor $-GM/c^2 r $  since one would get in this case
for the escape  velocity $ v^2   =  4GM/r$.  

Additional higher order terms in $v^2/c^2$ in the l.h.s. of
Eq.( \ref{fundament}) are therefore the only possible modification
of the free fall law for radial trajectories.
They  would result in first order corrections to Mercury's precession, which are
ruled out.  

In that respect the LLR experiments \cite{LLR} of the free falling Earth-Moon system
in the gravitational field of the Sun should also exclude such relativistic
corrections with higher accuracy.  Gravitational quantum interference experiments \cite{co} are very far from
providing possible additional information. 

The above arguments imply  that free fall is determined only
by $GM$ without kinetic and self energy corrections and  that
gravity does not contribute to the inertial mass $m_{o}$.
Thus all GR corrections to  Newtonian dynamics simply follows from the
treatment of {\it  free fall where 
 one mass is enough,  and cancels out in the motion in a given field}. 
This can be summarized as 
 
 $$ m_{I} = m_{0} = m_{0} (1-GM/c^2r  +v^2/2c^2)  =  m_{0} = m_{G}  $$
$I$ standing for inertial.

In conclusion our  treatment based on  Newton's law and  the free falling frames gets validated
beyond expectations and all the geometrical relations of GR are direct consequences.
Of course  the possible distinction between different substances  (WEP), ruled out by the
terrestrial experiments of E\H {o}t-Wash  \cite{eot}, is automatically implemented here. 

Let us finally comment on the role of moving frames which play such a fundamental
role in SR and GR, although sometimes with a misleading interpretation. 

In the former case they represent a physical entity:
e.g., in the the flying muon frame the atmosphere thickness is shorter
than the one measured on earth and the time needed to reach it is correspondingly
shorter. 
 
In the latter the accelerating falling frame is the basis for a construction
which describes gravity in terms of local ``inertial'' frames.
Thus the popular  expression that GR dilates spacial distance has no relation with
the physical fact that in SR distances for moving particles are shorter. 
That statement should be supplemented by the phrase:
in the Ss metric, whereas in the P-G it  does not happen
as also stressed by Mizony \cite {mizony}.
In that sense the P-G metric, with its clear and physically founded
combination of local SR with a global "Galilean" free fall
law (also preserving at infinity the gravity free absolute time),
has an interpretation which is "closer to reality".

\

\subsection{From the P-G to the Ss  metric }

\

In this paragraph the relation of the Ss metric   the P-G one  will be elucidated by
explicitly considering  the combination of Galilean transformations
for the free fall with the local space time Minkowski structure which yield the physical
time and length measured at rest.
In short:
\begin{equation} \label {GL}
  (dr,dt)_{global coord.}  \Longrightarrow _{Galilei}  \;  (dr',dt')_{EPIF}
    \Longrightarrow _{Lorentz Tranf.}  \;   (d\rho,d\tau)_{at rest}
    \end{equation}
i.e., on the basis of the globally defined coordinates $(r,t)$, extending the
Newtonian coordinates at infinity, EPIF differentials are given by eqs.(\ref{pseudogal}),
(eliminating gravity through free fall)
and then a SR transformation yields the frame at rest at a given point.  
Time and space coordinates at rest, $d\tau$, $d\rho$, are thus obtained  as   

\begin{equation}
  d\tau = \gamma(v) ( dt' + v dr') = \gamma(v) ( dt + v ( dr -v dt))
  = dt / \gamma(v)   + \gamma(v) v dr 
\end{equation}

\begin{equation}
d\rho = \gamma (dr'  + v  dt') = \gamma (dr -vdt + vdt) =  \gamma  dr
\end{equation}
where 
$ \gamma(v) = \gamma[v(r)] =\frac{1}{\sqrt{1-v^2/c^2}} = \frac{1}{\sqrt{1- 2GM/c^2r}} \, $.

We stress the role of the Galilei transformation in the above derivation.
The last  equation shows
a dilation of lengths, contrary to the ordinary SR effect.

According to the second to last, times are indeed shortened,
as in SR, were it not for a space dependent term.   
Since that term {\it  does not alter the time rate at a given space point,
  it can be dropped through a redefinition of the global time}: 


\begin{equation} \label {tS}
  dt \to  dt +  v/c^2 /(1-v^2/c^2) dr = \gamma d\tau  \equiv dt_{S}
  \end{equation}
  
Together with $dr_S \equiv dr$, this gives the Schwarzschild coordinates
and the Schwarschild metric

\begin{equation}
  ds^2  =  c^2 d\tau^2 - d\rho^2 = (1-v^2/c^2)  c^2  dt_{S} ^2 - dr_{S} ^2/(1-v^2/c^2) -
  r^2 d\Omega^2
\end{equation}


The difference between the two notions of time is particularly  evident
in the treatment of light deflection, where in the P-G metric a linear
factor in the velocity shows up, with a presumed huge effect when observed
half way (e.g. on earth in the measurement of parallaxes).
The point is that one must not confuse the P-G notion of simultaneity
with the  one defined by clocks at rest 
(and therefore on earth apart from an easy relativistic correction for its motion).
 
It should also be noticed that the Ss metric has  spurious singularities
not only at the S radius $r = 2GM/c^2$, as well known,  but also at
$ r = \infty$ where the Newton time $t$ and $t_S$ differ by $\simeq  \sqrt{r}$.

{\it In conclusion  the P-G coordinates have the advantage of the underlined
  physical foundation, the lack of singularities, no necessity of an equation
  of motion beyond Newton's law  and can be directly and simply applied to all
  processes}, apart from the discussion of equal time geometrical effects, as
the parallax, where the Ss coordinates give a notion of simultaneity which
coincides with the one at rest.
 
Let us recall that the requirement to eliminate the off diagonal term of the
P-G metric is  generally accomplished just by redefining time in an ad hoc way,
as in Eq.(\ref{tS}), without any discussion about its physical meaning,
nor about its effects in the interpretation of experiments.  


  Finally, let us underline  an inherent "paradox" of GR.
The pretension that coordinate independence of the formulation is fundamental backfires,
  in the sense that Newton's absolute time  not only has the right of citizenship, but 
  gives rise to an independent description based on fundamental
  physical motivations. 
 
\

\subsection  {\bf On alternative derivations}

\

Ever since the appearance of GR, the endeavor to find other solutions 
than Schwarz\-schild's, to "derive" it from SR and to eventually propose
alternative theories has been paramount.   

To start with, let us recall that Einstein's rebuttal of the Painlev\'{e}-Gullstrand
\underline{solution} has led to an ostracism (their  metric is not even mentioned
in most textbooks) which has lasted till almost the end of the last century.
Only recently the P-G metric has been reevaluated as a singularity-free solution.
In addition, it has been realized that it could be obtained directly from basic principles ,
without recursion to GR.

This possibility has provoked a heap of warnings: that it could be only heuristic,
that it only may apply to the weak field case, accompanied as well
by the (trivial) argument that it cannot reproduce all of GR results.
In connection with the first points  we  want to comment
on some of the most relevant and cited articles. 

Schiff's \cite{schiff} work had already been criticized by Schild  \cite{schild}.   
The usual result for time had  been obtained by using a
SR  argument, comparing local time with that of gravity free infinity
via a flying, time-shortened, clock. However his (incomplete) argument about
space cannot be correct since in the end, contrary to SR, the velocity of light
is not constant nor isotropic.
His statement about Mercury's perihelion being {\it the} crucial test of
GR has already been commented upon above.

Kassner's \cite {kass}  work is relevant in the present context because of his
discussion of the necessity of supplementary assumptions in order
to derive the Ss metric on the basis of pre-general-relativistic physics
{\it alone}, i.e., SR, the Einstein EP and the "Newtonian limit".

This is not contradictory with our findings. As a matter of fact,
the Newton law is used by us globally, not only to first order at 
infinity, supplemented by the two (almost unescapable because of our motivations)
subsidiary conditions on space (length of the circumference) and absolute time.   
They can be seen as substitutes of Kassner's two additional ``postulates'',
which serve the same aim but which are, in our opinion, less transparent
and motivated.  

Czerniawsky's point of view \cite{cern} is the closest to ours. 
Our assumption on the Euclidean properties of space, at equal ``free fall'' times,
is somewhat hidden in  his considerations about the EP. As a result,
his treatment does not include the dependence
on {\it two} functions of the radius (eqs.(1),(2)),
a general fact already recognized in \cite{gru}.
On the other hand we agree with Czerniawsky's considerations on the
difference in the notion of simultaneity between the Ss and PG metrics
and on the physical significance of the time-reversed PG metric.

Finally Visser \cite{vis} and Padmanabhan  \cite {padma}
have strived to maintain the inadequacy of the free fall approach
for the following reasons: to be only a weak field approximation of a
more general theory and to be heuristic since it does not reproduce 
the Kerr metric.  
The first point has already been  commented upon.
The second is irrelevant in the present context. For  rotating masses results  
have  been reproduced successfully via gravitomagnetism
in a parameter free way just from SR without the need to invoke GR
\cite {christbara}.
 
In general, it is important to underline the peculiarity of proper
time effects, with respect to those involving space. is is clear already to first order, since time contraction can be obtained as
a SR effect, while the treatment of space depends
on the overall analysis, see eq.(\ref {GL}). 

Their effect to first order has been evaluated by  Einstein  using only mass energy equivalence and is as  follows:

Consider an atom at  $B = r' = r +h$ and an identical one at $A=r$.
Then the photon emitted by B reaches A,
because of the coupling between its energy and the gravitational field,
with a greater energy due to the effect of the gravitational field.
The photon  frequencies at the two places are related by

\begin{equation}\label {fundamental}
  \hbar \omega (1 - GM/c^2 r) =   \hbar \omega' (1 - GM/c^2 r')
\end{equation}

from which it trivially follows 

\begin{equation}\label{om1}
\omega   = \omega '  \frac{{ 1 - GM/c^2r'}}{ { 1 - GM/c^2r}} \simeq  \omega' (1+ gh/c^2) 
\end{equation}

This implies the reverse relation for times 
 
 \begin{equation}\label{om2}
t'   = t   \frac{{ 1 - GM/c^2r'}}{ { 1 - GM/c^2r}} 
\end{equation}
 i.e. that time runs quicker in regions of smaller gravitational field.
 When the comparison is made with respect to $\infty$, where gravity is absent,
 one gets the proper time at r denoted by $\tau$,

 \begin{equation}\label{om3}
t'   = t_{\infty} = \tau  / ( 1 - GM/c^2r) 
\end{equation}
 and this agrees to first order with the above result
 from the invariant interval. 

 {\it Notice that a basic form of principle of equivalence has been tacitly assumed:
   atoms are the same (as locally measured) in different points
   of a gravitational field.}
   Otherwise a correction factor would arise.

 This goes along with the parallel argument about atomic energy levels.
 The  mass $m$ {\bf at rest} in a gravitational field of $M$
 at the height $R_{T}$ has an energy 
\begin{equation}
E_{0} = m_{0}c^2 ( 1  - GM/c^2 R_{T})
\end{equation}
and at $R_{T} +h$
\begin{equation}
E' = m_{0}c^2 ( 1  - GM/c^2 (R_{T}+h))
\end{equation}
It follows that at the earth surface 
\begin{equation}
E' - E_{0} \simeq m_{0}c^2 GM/c^2 R_{T}^2)h = m_{0} g h
\end{equation}
This energy difference exactly corresponds in classical terms
to the gravitational potential energy difference or, in other words,
to the work done against the standard Newtonian force
$ \bf F = m \bf g$.
The two arguments are consistent because of {\it local} energy conservation
of the atom-photon systems.
They also show that the use of the gravitational interaction energy is consistent,
to first order, with the dynamical treatment based on the elimination of gravity
in free fall motion.

Independently of the first order approximation, the peculiarity of pure time effects
is that they are physical, i.e. coordinate independent.
In fact, time effects are given by the invariant interval at {\it a fixed point in space},
and, as observed above,  such a notion is independent of coordinate transformations preserving
stationarity of the metric tensor.

This does {\emph not} happen (not even to first order)
for the other components of the metric tensor,
which are in fact different in the Ss and PG formulation.



\section  {Rotating frames and the Sagnac effect }

\

We pass now to a subject, not directly related to gravitation, whose treatment
may  help in shedding some more light on the use of metrics and of synchronization. 

The Sagnac effect has a long history,  remarkable practical applications and has caused a
considerable amount of discussions concerning its connection with SR and GR.

In its standard form two counter propagating photon beams in a circular waveguide mounted
on a disk are made to interfere after having traveled one circumference. 
When the disk is put in rotation with angular velocity $\Omega$,
the interference figure is seen to shift by an amount proportional to $\Omega$.

Let us consider the problem from the point of view of the external observer
(inertial frame). For
him, light propagates of course with velocity $c$; however  when the disk rotates  the interference of the two 
light waves is observed
at a moving angle, $\theta = \Omega t $.

The  lenghts $ l_{1,2}$ of the two paths satisfy 

$$ l_1 - l_2 = 2 \theta r \, ,$$
where $\theta$ is the shift of the angle in the traveling time  and 
$r$ the radius of the disk (with time and distances measured in the fixed frame).
This corresponds, for light of frequency  $\omega$, to a shift in phase
$$ \Delta \varphi = 2 \theta r \, \omega / c  $$
To first order in $\Omega$, the traveling time of the two rays is $ t \simeq 2 \pi r /c $ and
therefore 
\begin{equation} \label {delt}
  \Delta \varphi  \simeq \frac {4 \pi r^2 \omega \Omega }{ c^2 }
  = \frac{ 4 \omega \Omega S }{c^2} \,  ,
 \end{equation} 
$S$ standing for the area perpendicular to the rotation
axis enclosed by the given contour.

This is all,  since the effect  is frame independent. 
However for the sake of the argument and in order to make contact with the previous
treatment of gravity, let us consider it from another point of view.

\
 
{\bf    On  the invariant interval in rotating frames }

\

The above kinematical constraint about the meeting of two rays at a moving point
can of course be written as the condition of meeting at {\it the same point
 in a rotating coordinate system}
and can be therefore discussed in terms of light propagation {\it in such coordinates.}
This does {\it not} mean that quantities
{\it measured} `` on a rotating body'' enter the discussion and in fact the
introduction of such ``physical frames'', in particular of local frames
associated to observers at each point on the disk, is not necessary.

Consider then a uniformly rotating reference  system, whose local cylindrical coordinates
are denoted by $(t,r, z, \phi_R )$ connected to those of the fixed inertial one
$(t,r,z, \phi)$ by 
\begin{equation} \label{rota}
\phi_R = \phi  -  \Omega \; t \, ;
\end{equation}
the invariant interval in the rotating system reads

\begin{equation}
      ds^2  =  c^2  \,dt^2 - (r d\phi_{R} + \Omega r dt)^2 - dz^2 - dr^2  \, ,
\end{equation} 
in complete analogy with gravity in our Newtonian coordinates.
This immediately yields that light propagates tangentially with velocity  
$$ c_{\perp} = \pm c - \Omega r \, , $$
as in the  composition of light velocity with that of the free fall frame in gravity.

The previous equation can be rewritten in terms of $ v = \Omega r$ and $ dy = r d \phi_R $  
\begin{equation}\label{dsrot}
      ds^2  =  c^2( 1  - v^2/c^2 ) \,dt^2 - 2 v dt  dy  -   \; dy^2 - dr^2   - dz^2 
\end{equation}
The similarity with the P-G formula is once more apparent.
The essential difference is that now $v$ is independent of $y$,
and in fact the differential $r d\phi = dy + v \, dt$  \;  from Eq.(\ref {rota}),
corresponding to the EPIF differential  $ dr' = dr - v(r)\, dt $,  is now exact.

The above difference between light velocities in the two directions easily leads 
to the same result as before.
We emphasize that the analysis applies to first order in $\Omega$,
that no relativistic effect
appear to that order and that the above discussion of the relativistic interval
has nothing to do with Lorentz transformations,
rather expressing the interval in the inertial frame
in terms of different coordinates (intervals in such coordinates
coinciding with  measured intervals ``on the moving disk'' only
in the non-relativistic limit).

It is also of some interest to write the above relativistic interval in
``Schwarzschild coordinates'': the off-diagonal term can be disposed of
along the previous lines via the transformation
\begin{equation} \label {SSagn}
\begin{split}
& dt_S = dt + v/(1-v^2/c^2)  dy \\
 & dy_S = dy 
\end{split}
\end{equation}
and for the relevant part (i.e. apart from $dr^2$ and $dz^2$ terms)
the invariant interval takes the ``Schwarzschild form'' 
\begin{equation}
      ds^2  =  c^2( 1  - v^2/c^2 ) \, dt^2_S  \;     -  dy_S^2/(1 - v^2/c^2)  
\end{equation}
In both forms, the invariant interval is {\bf not}  Minkovskian, and in fact light
velocity is different from $c$ both in the "P-G" and in the ``Schwarzschild''  coordinates,
where the tangential velocity is direction independent:
$$ c_S =  c \; (1 -  v^2/c^2) $$




 
 



This is compatible with the Sagnac effect because
Eq.(\ref{SSagn}) only gives rise to a {\it local} notion of time and global, topological,
effects are hidden in the angular nature of the $y$ variable. 
In Ss coordinates the Sagnac effect comes in fact from the difference in time
coordinates obtained after following closed paths.
The time difference between two path enclosing the circle in opposite directions is
$$\Delta t_S = 2/c^2 \oint \frac{ \Omega r^2 }{(1- \Omega ^2 r^2/c^2)} \, d\phi $$
which for low angular velocities yields 
$$\Delta T = \frac{ 4 \Omega S}{ c^2 }$$ 
corresponding to the result obtained via elementary considerations
at the beginning of the paragraph.

\

\

\section{Conclusions}

\

{\it In the present work  Einstein's equations have been shown  to be
  unnecessary in the static symmetric case, where all GR results have been
  obtained via EP, Newton's law and local SR.}th
The result is the PG metric,  which realizes  the Einstein program
of ``eliminating gravity locally'' through free fall motion.

The  calculational  ingredient has been the usual  variational principle 
(like the Fermat one for light)  applied to the infinitesimal invariant non Minkovskian interval and the associated  Euler-Lagrange equations realizing what usually appears as the result of   a rather complicated and formal description of gravitation.  
 
The connection with the Ss  GR  solution,   which is superfluous  given the explicit calculations of the present approach  (and which had been considered  as a necessary endorsement for the Painlev\'{e}-Gullstrand solution)  only helps in clarifying  the arbitrariness  of attaching physical significance to the metric. 
In particula,  space-time curvature is perfectly compatible  with  Newtonian absolute time and Euclidean space.

The treatment  of rotating  frames, which played a role in the genesis of GR,
leads, at each point, to a metric of the same form, with the same role of local frames as
in the case of gravitational free fall. The only difference is here the existence of a global Minkovski frame.

\

{\bf  Acknowlegments}

\

We would like to thank Dr.  E.  Cataldo for having brought to our attention a relevant piece of literature,  prof.s F.  Strocchi and  P. Menotti for many discussions  and Dr. A. Rucci for the drawings.

\


\begin{thebibliography}{999}
 
 \bibitem{newton}
I. Newton,  Philosophiae Naturalis Principia Mathematica (ÒMathematical Principles of Natural PhilosophyÓ), London, 1687; Cambridge, 1713; London, 1726. 

 
 \bibitem{einstein}
A.Einstein, Die Grundlage der allgemeinen Relativit\"atstheorie ,   Annalen der Physik. 49, 1916, S. 769Ð822. 
 
\bibitem {schiff}
 L.I.  Schiff,  Am. J. Phys. 28 (1960) 340.


 
\bibitem{cern}
Jan Czerniawski,arxiv.org/abs/gr-qc/0201037 (2004) 


				arXiv:gr-qc/0611104 (2016)
 

 \bibitem{gru}
R.P. Gruber et al., Am. J. Phys. 56, 265  (1988)
 


   
\bibitem{vis}
M. Visser arXiv : gr-qc/0309072v3 2004


\bibitem{padma}
T. Padmanabhan,  link.springer.com/article/10.1007/s12045-009-0101-x 


\bibitem{kass}
    K. Kassner, Eur. J. Phys. 36 (2015) 065031 ,   arXiv:1502.00149 [pdf, ps, other]



\bibitem{schild}
A.Schild, Am.J. Phys. {\bf 28}  778, (1960) 

\bibitem{rind} 
W. Rindler, Am. J. Phys. {\mathbf 36}, 240-245 (1968)

\bibitem{traut}
N. Strauman, arXiv:astro-ph/0006423

\bibitem{el}
J. Ehlers and W. Rindler, Gen. Rel. Grav. 29, 519 (1997

\bibitem{pirani}
Pirani, F.A.E. , Acta Phys. Polonica, 15, 389 (1956) 

\bibitem{fey}
R. Feynman, Lectures of gravitation, Penguin Books

\bibitem{pain}
P. Painlev\'e, ÒLa m\'ecanique classique et la theorie de la relativit\'eÓ, C. R. Acad. Sci. (Paris) 173
677-680 (1921).

\bibitem{gull}
A. Gullstrand, ÒAllgemeine l\"osung des statischen Eink\"orper-Problems in der Einsteinschen
GravitationstheorieÓ, Arkiv. Mat. Astron. Fys. 16(8) 1Ð15 (1922).

\bibitem{ss}
  Schwarzschild, K. (1916). "\"Uber das Gravitationsfeld eines Massenpunktes
  nach der Einsteinschen Theorie". Sitzungsberichte der K\"oniglich Preussischen Akademie der Wissenschaften. 7: 189Ð196.


\bibitem{berk}
Berkeley Physics Course, Mechanics, McGraw-Hill (1962)


\bibitem{mond}
  Milgrom, M. (1983),  Astrophysical Journal. 270: 365Ð370.

 \bibitem{damour}
 T. Damour,  La Matematica, vol. IV, G. Einaudi Editore,  453 (2010)
 


 
 \bibitem{LLR}
 J.G.Williams, S.G.Turyshev and D.H.Boggs, arXiv : gr-qc/0411113v2 (2004)  ,   Int. J. Mod. Phys. D 18, 1129 (2009). https://doi.org/10.1142/S021827180901500X

  
 \bibitem{co}
R. Colella, A.W. Overhauser and S.A. Werner. Phys. Rev. Lett.,  34  (1975), p. 1472.

\bibitem{eot}
 E\"ot-Wash Group,  Class. Quantum Grav. 29, 184002 (2012)

\bibitem {mizony}
M.  Mizony,  math.univ-lyon1.fr/~mizony/cargese1
 



 \bibitem{christbara}
 P. Christillin and L.Barattini , Gravitomagnetic forces and quadrupole gravitational radiation from special relativity, arXiv:1206.4593 (2013) 



\end{thebibliography}
\end{document}